# A 'Critical' Insight into Pretransitional Behavior and Dielectric Tunability of Relaxor Ceramics

**Sylwester J. Rzoska[1*], Aleksandra Drozd-Rzoska[1*], Weronika Bulejak[2*], Joanna Łoś[1], Szymon Starzonek[3], Mikołaj Szafran[2], Feng Gao[4]**

[1]Institute of High Pressure Physics Polish Academy of Sciences, ul. Sokołowska 29/37, 01-142 Warsaw, Poland
[2]Faculty of Chemistry, Warsaw University of Technology, Noakowskiego 3, 00-664 Warsaw, Poland
[3]Laboratory of Physics, Faculty of Electrical Engineering, University of Ljubljana, Ljubljana, Slovenia
[4]State Key Lab. of Solidification Processing, School of Materials Science and Engineering, Northwestern Polytechnical University, 710072 Xi'an, Shaanxi, PR China

(*) S. J. Rzoska, e-mail: sylwester.rzoska@unipress.waw.pl
(*) A. Drozd-Rzoska, e-mail: arzoska@unipress.waw.pl
(*) W. Bulejak, e-mail: weronika.bulejak.dokt@pw.edu.pl

**Abstract**

The model discussion focused on links between the unique properties of relaxor ceramics and the basics of Critical Phenomena Physics and Glass Transition Physics is presented. It indicates the significance of uniaxiality for appearing mean-field type features near paraelectric – ferroelectric transition. Pretransitional fluctuations, increasing up to grain size and leading to inter-grain, random, local electric fields, are indicated to be responsible for relaxor ceramics characeristics. Their impacts yield the pseudospinodal behavior associated with 'weakly discontinuous' local phase transitions. The emerging model re-defines the meaning of the Burns temperature and polar nanoregions (PNRs). It offers a coherent explanation of 'dielectric constant' changes with the 'diffused maximum' near paraelectric – ferroelectric transition, the sensitivity even to moderate electric fields (tunability), and the 'glassy' dynamics.

These considerations are confronted with experimental results for the complex dielectric permittivity studies in $Ba_{0.65}Sr_{0.35}TiO_3$ relaxor ceramic, covering ca. 200K range, from the paraelectric to the 'deep' ferroelectric phase. The distortions-sensitive and derivative-based analysis revealed the preference for the exponential scaling pattern for $\varepsilon(T)$ changes in the paraelectric phase and in the surrounding of the paraelectric-ferroelectric transition. It may suggest the Griffith-phase type behavior, associated with the mean-field criticality disturbed by random local impacts. The discussion of experimental results is supplemented by relaxation times changes and the coupled energy losses analysis. The studies also led to the description of tunability temperature changes with scaling relations.

# 1. Introduction

Relaxor ceramics remain a cognitive challenge despite seven decades of studies [1-26], and the significance for innovative applications: from varactors, signal tunable filters, phase shifters, and frequency-selective surfaces significant for conformal antennas, to possible electrocaloric effect implementations [26-31]. Particularly important are unique dielectric properties and their sensitivity to the external electric field, the tunability. The significance of relaxor ceramics shows the growth in research reports number since 2020: 24% rise in 2022 and 70% rise (up to about 3600 papers) is expected in 2023 [32]. As for applications the global market is expected to quadruple between 2022 and 2029 to around USD 16 billion [33].

Unique properties of relaxor ceramics are mainly related to dielectric constant changes near the paraelectric – ferroelectric transition. For the reference case of 'homogeneous', canonic ferroelectric materials dielectric constant is portrayed by the Curie-Weiss (CW) relation [1-25, 34-38]:

$$\varepsilon(T) = \frac{A_{CW}}{T - T_C} \quad (1a) \qquad \Rightarrow \qquad \frac{1}{\varepsilon(T)} = A_{CW}^{-1} T - A_{CW}^{-1} T_C \quad (1b)$$

where $A_{CW} = const$ and $T_C$ is the Curie-Weiss critical temperature;

For inherently 'heterogeneous' relaxor ceramics, instead of the 'infinite' singularity $\varepsilon(T \to T_C) \to \infty$ (Eq. (1)), a 'diffused' in temperature maximum of $\varepsilon(T)$ appears [1-25]. The next unique feature is related to strong changes of dielectric constant when applying even a moderate external electric field. It is described by so-called tunability [2, 8, 10, 13, 20, 24]:

$$T(\%) = \frac{\varepsilon(E \to 0) - \varepsilon(E)}{\varepsilon(E \to 0)} \times 100\% \qquad (2)$$

Finally, dynamics of relaxor ceramics exhibist scaling patterns known for glass-forming systems in the previtreous domain. The hallmark is the super-Arrhenius (SA) temperature evolution of the primary relaxation time, for which the Vogel-Fulcher-Tammann (VFT) is used as the main replacement equation [7, 14, 15, 17-20]:

$$\tau(T) = \tau_\infty \exp\frac{E_a(T)}{RT} \quad (3a) \qquad \Rightarrow \qquad \tau(T) = \tau_\infty \exp\frac{D}{T - T_0} = \tau_\infty \exp\frac{D_T T_0}{T - T_0} \quad (3b)$$

The right part (Eq. 3a) is for the canonic SA relation, with the apparent (temperature-dependent) activation energy $E_a(T)$. It simplifies to the basic Arrhenius equation for $E_a(T) = E_a = const$, in the given temperature domain; $R$ denotes the gas constant.

For the VFT model- equation: $E_a(T) = Dt = (RD_T T_0)t^{-1}$, and $t = (T - T_0)/T$ for the relative distance from the extrapolated singular VFT temperature $T_0$ [39]. In glass-forming systems $T_0$ is located below the glass temperature $T_g$, by 'convention' linked to $\tau(T_g) = 100s$. The amplitude $D = const$; $D_T$ is called the fragility strength [39].

For determining mentioned properties essential meaning has broadband dielectric spectroscopy (BDS), which output results can be presented as the complex dielectric permittivity: $\varepsilon^*(f, T) = \varepsilon'(f, T) - i\varepsilon''(f, T)$. The real part enables determining the canonic dielectric constant. It is associated with the so-called static domain of $\varepsilon'(f, T = const)$ spectrum, where a frequency shift does not change significantly its value. For dipolar dielectrics, it is located within $1 kHz < f < 10 MHz$ range. For lower frequencies (LF), below the static domain, the strong rise of both $\varepsilon'(f)$ and $\varepsilon''(f)$ occurs. It is linked to the impact of ionic contaminations. The response related to relaxation processes appears for higher frequencies above the static domain [39].

In relaxor systems, the temperature evolution of dielectric constant: $\varepsilon'(T, f = const)$, near paraelectric – ferroelectric transition manifested as the diffused maximum and CW (Eq.(1)) described 'branches' detected for a set of scanned frequencies. Parameters describing the maximum, $(\varepsilon'_{max}, \varepsilon_m)$ and $(T_{max}, T_m)$, are frequency-dependent [3-19, 24, 25]. It indicates that for relaxors, one should consider the real part of dielectric permittivity rather than the canonic dielectric constant.

Regarding dynamics, significant is the primary loss curve $\varepsilon''(f, T = const)$ characterizing the relaxation process associated with permanent dipole moments. Its time-scale estimates the peak frequency, namely the primary relaxation time: $\tau = 1/\omega_{peak} = 1/2\pi f_{peak}$ [39]. By tradition, for ferroelectric systems, including relaxors, temperature scans for subsequent frequencies are often carried out, leading to the manifestation of primary loss curves in $\varepsilon''(f = const, T)$ dependencies. In such a case, experimental data are portrayed via the following relation [1-25]:

$$f(T) = f(T_m) = f_\infty exp\frac{E'_a(T)}{T} = f_\infty exp\frac{D'}{T_m - T_0} \tag{4}$$

where $T_m$ is for temperatures for the maximum of $\varepsilon''(f = const, T)$ detected for temperature scans using subsequent frequencies; $T_m = T_{max}$ is the temperature describing the loss curve maximum for the scan carried out for the given frequency $f$.

Eq. (3) converts into Eq. (4) for $\tau(T) \to 1/f(T)$ and $T \to T_m$.

The 'glassy, previtreous' dynamics is also associated with the non-Debye, multi-time, distribution of relaxation times. It manifests via the 'broadening' of the primary loss curve above the single-relaxation time Debye pattern. Most often it is portrayed via the Havriliak-Negami (HN) relation, commonly used also for relaxor ceramics [3-5, 9, 12, 13, 15, 19, 22, 24, 25, 39]:

$$\varepsilon^*(f) = \varepsilon_\infty + \frac{\Delta\varepsilon}{(1+(i\omega\tau)^a)^b} \tag{5}$$

where power exponent $0 < a, b < 1$

for $a, b = 1$ Eq. (5) is simplified to the basic Debye equation associated with a single relaxation time. In Eq. (5) $\Delta\varepsilon = \varepsilon - \varepsilon_\infty$ is called dielectric strength, and describes the dipolar contribution to the 'total' value of the dielectric constant; $\varepsilon_\infty$ is the non-dipolar permittivity related to electronic and atomic contributions.

Studies in supercooled glass-forming liquids showed that power exponents in Eq. (5) can be used as metrics for the distribution of primary relaxation times, which is well shown by the link to Jonsher scaling of primary loss curves $\varepsilon''(f, T = const)$ [40-44]:

$$\frac{\varepsilon''(f<f_{max})}{\varepsilon''_{max}} = a'f^m \quad \Rightarrow \quad log_{10}(\varepsilon''(f)/\varepsilon''_{max}) = log_{10}a' + m log_{10}f \tag{6a}$$

$$\frac{\varepsilon''(f>f_{max})}{\varepsilon''_{max}} = b'f^{-n} \Rightarrow log_{10}(\varepsilon''(f)/\varepsilon''_{max}) = log_{10}b' - nlog_{10}f \qquad (6b)$$

where $T = const$, and $a', b' = const$

The following link between the distribution metric for HN Eq. (5) and Jonsher Eq. (6) takes place: $m = a$ and $n = ab$. The reference Debye relaxation is related to $m = n = 1$.

Notable that the analysis based in Eqs. (6a) and (6b) enables the reliable determing of the relaxation time, using the condition: $dlog_{10}(\varepsilon''(f))/dlog_{10}f = 0$, for $f = f_{peak}$ and $\tau = 1/2\pi f_{peak}$. Alternatively, the relaxation time is determined using the HN Eq. (5), leading to five-parameters nonlinear fitting.

Notable that for glass forming systems, the SA (Eq. (3)) and non-Debye (Eq. (5)) behavior takes place on cooling from the ultraviscous/ultraslowed domain to the amorphous solid glass. It is associated with the time-scale $\tau(T_g) \sim 100s$ [39]. For relaxor systems the transition is associated with the 'diffused' paraelectric – ferroelectric transition and the mentioned time scale is not reached [2, 8, 10, 13, 20, 24]. It should be stressed that such complex dynamics is absent for basic 'homogeneous' ferroelectric systems.

In relaxor ceramics, the temperature at which the distortion from the CW behavior (Eq. (1)) occurs on approaching the transition is called the Burns temperature [3, 4]. It is linked to the onset of Polar Nanoregions (PNRs), a key concept used for explaining unique relaxors' features [5-20, 22-26]. It is stated that the emergence of PNRs begins to form rapidly through the interaction among adjacent dipoles and orients between the states with the same energy and contributes less to the dielectric permittivity because of violent thermal fluctuation. The enhanced interactions among the dipole clusters increase the correlation length, giving the PNRs local field properties. The PNRs can be reoriented under the effect of the electric field, significantly changing the dielectric permittivity characterized by the mentioned deviation from Curie–Weiss law. Following such a picture, the 'microscopic fluctuations' models in which local fluctuations related to PNRs cause local changes in the Curie temperature $T_C$ was introduced [5-19, 24-26, 45, 46]. Assuming the Gaussian-type distribution of $T_C$ the following relation for portraying dielectric constant changes was proposed by Uchino and Nomura [2, 15]:

$$\frac{1}{\varepsilon(T)} = \frac{1}{\varepsilon_m} exp\left(-\frac{(T-T_A)^2}{2\sigma^2}\right) \approx \frac{1}{\varepsilon_m}\left[1 + \frac{(T-T_A)^2}{2\sigma^2} + \cdots\right] \qquad (7)$$

for $T > T_m$, $\varepsilon_m = \varepsilon_{max}$. Uchino and Nomura proposed to assume $T_A = T_m$ and generalize the above relation to arbitrary power exponent $1 \leq \gamma \leq 2$, what led to the commonly used semi-empirical relation [15]:

$$\frac{1}{\varepsilon(T)} - \frac{1}{\varepsilon_m} = C'^{-1}(T-T_m)^{\gamma'} \qquad (8)$$

It can portray experimental data even for $T - T_m > 1 \div 3K$ [2, 5, 6, 12, 13, 15, 18, 19, 22-24] It is stressed that for $\gamma = 1$, Eq. (8) 'reduces' to CW Eq. (1), which has to lead to the conclusion that $\gamma' \to 1$ is coupled to $T_m \to T_C$ and $\varepsilon_m \to \infty$. Nevertheless, the link of the exponent $\gamma'$ in Eq. (8) to well-defined critical exponents [15, 47] is not clear, in the opinion of the authors.

Notable that in ref. [6], a different form of dielectric constant changes in the surrounding of paraelectric-ferroelectric transition via the model considering the impact of relaxation polarization processes associated with PNRs led to two contributions originating from the thermally activated flips of the polar regions, and the second one represents the contribution from the 'other' polarization process. It led to the following relation for $T > T_m$ :

$$\varepsilon(T) = \varepsilon_\infty + \varepsilon_{ref.} exp(a - bT) \qquad (9)$$

where in the given case parameters $a, b = const$, and coefficient $b$ is related to the product rate of PNR in the material

The behavior in the ferroelectric state, for $T < T_m$, was also derived [6]:

$$\varepsilon(T) = \varepsilon_\infty + A(T)(ln\omega_0 - ln\omega) \qquad (10)$$

where $\omega_0$ is the average relaxation frequency of a polar unit cell that is independent of the temperature, i.e., $ln\omega_0 = const$, $A(T)$ is an intrinsic parameter of the relaxor material.

So far, experimental results for relaxor systems are commonly scaled via Eq. (7) or its parallels. The reorientation of PNRs, characterized by the relaxation time (τ), is also the reference for models focusing on non-Arrhenius behavior of the primary relaxation time, for which the VFT relation is used as the scaling reference. In the opinion of the authors, a problem appears when taking into account that PNRs are related to the Burns temperature, the onset of the distortion from the CW behavior on cooling towards paraelectric – ferroelectric transition, whereas the glassy dynamics portrayed by Eq. (3) is observed on both sides of $T_B$ [5, 9, 12, 13, 15, 18, 19, 22-26, 28, 45, 46].

Despite decades of studies, the ultimately and commonly accepted model explaining mentioned features of relaxor ceramics remains lacking [6-8, 10, 11, 14, 15, 17, 20, 26]. The combination of 'glassy' dynamics and 'distorted critical-like behavior (Eq. (1)) properties still is a challenge. The problem constitutes even the coherent addressing canonic features mentioned above and deriving check-point canonic relations. Simple and fundamentally justified scaling dependences supporting modeling can be particularly significant for supporting the expected boost in relaxors- based innovative devices [32-35].

In this work, we propose to look at the debatable above properties of relaxor materials from a slightly different perspective than before, namely with an explicit reference to the foundations of *Critical Phenomena & Phase Transitions Physics* [47-49] and *Glass Transition Physics* [39, 50, 51] and then to confront the emerging conclusions with existing and new experimental results, based on research carried out specifically for this work. They also were used to search for further, hitherto unaddressed, experimental characterizations of a given phenomenon.

## 2. Materials & Methods

BST sample was prepared using $BaCO_3$ (>98%, Chempur, Poland), $SrCO_3$ (>98%, Chempur, Poland), and $TiO_2$ rutile (>99.9%, Sigma-Aldrich). Materials in stoichiometric proportions ($Ba_{0.65}Sr_{0.35}TiO_3$), were ball-milled for 7 h in water and ethanol, subsequently dried and calcined at 1050°C for 2 h, and finally, barium strontium titanate was synthesized in the high-temperature solid-state reaction carried out at 1340°C for 2 h. The sintered material was ground with water and zirconia grinding media on a Witeg BML-6 ball mill at a speed of 300 rpm for 7 hours, and after drying, the samples of 20 mm diameter and 5 mm thick were obtained by die pressing and sintered at 1300°C for 1 h. The densities of the samples were measured using a helium pycnometer AccuPyc II 1340 (Micromeritics).

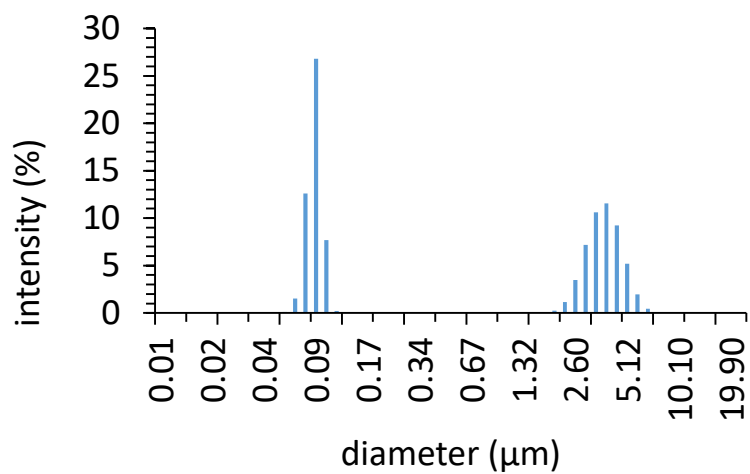

**Figure 1.** Results of particle size distribution analysis

The density of the synthesized powder was $5.629 \pm 0.004$ g/cm$^3$, while the density of the sintered sample was $5.612 \pm 0.005$ g/cm$^3$. The average particle size measurements were performed using the Laser Scattering Particle Size Distribution Analyzer LA - 950 by HORIBA. Figure 1 shows the particle size distribution of synthesized barium strontium titanate powder. The average particle size was 1.88 μm.

The powder consists of two fractions: small particles (0.06 - 0.13 *μm*) and larger agglomerates (2-7 *μm*), which were probably formed by re-aggregation of small particles during the milling process. Powders X-ray diffraction patterns were recorded at room temperature on Panalytical X'PERT PRO MPD X-ray diffractometer with a Cu anode. An X-ray diffractogram was made in the angular 2θ range from 5 to 81° for the powder sample and used to identify the phase composition. It was quantitatively analyzed using the Rietveld method, also employed for calculating the size of crystallites.

A sample holder with a spinner was used in this study. The size of crystallites and lattice distortions were determined directly from the Sherrer equation for 110 BST reflex. The coarse-crystalline calcite of natural origin and its reflex 104 were used as a half-width standard for the measuring system. The unit cell parameters were refined by the Rietveld method in quantitative analysis. The results of XRD qualitative and quantitative analysis are shown in Table 1. The synthesized BST consisted of a high (99.3%) percentage of BST in the assumed stoichiometry (Ba$_{0.65}$Sr$_{0.35}$TiO$_3$) in cubic (77.1%) and tetragonal (22.9%) phases and a small (below 1%) addition of cubic BaTiO$_3$.

**Table I** The composition, the structure, and the size of crystallites for the tested relaxor ceramic.

| Composition | Share (%) | Crystalline Structure type | The share of the given CS type % |
|---|---|---|---|
| $Ba_{0.65}Sr_{0.35}TiO_3$ | 99.3 | cubic | 77.1 |
|  |  | tetragonal | 22.9 |
| $BaTiO_3$ | 0.7 | cubic | 100 |

Based on the microstructure observations of the sintered sample performed by using a scanning electron microscope (Fig. 2), the grains grew approximately five times larger. During the sintering process of the agglomerates, pores and grain boundaries disappear so that we can observe in the sample structure sintered agglomerates, with a size in the range of 2-10 micrometers. The visible defects in the sample were probably caused by the grains being torn out while breaking the sample for observations.

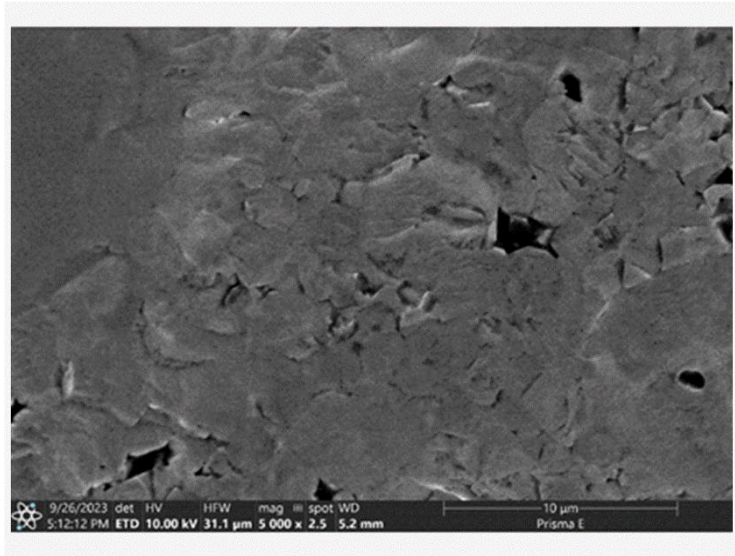

**Figure 2.** Scanning electron microscope picture of the sintered tested ceramic sample

Ceramic samples were then cut into 1 mm thick disks for Broadband Dielectric Spectroscopy (BDS) studies [51]. They were carried out using Novocontrol BDS Alpha spectrometer, enabling high-resolution studies up to 5 – 6 digits permanent resolution in broad frequency and temperature ranges. The coupled Quattro system controlled the latter. The adjustment of the system elements made by the manufacturer allows the removal of all parasitic capacitances and registration directly in the ensemble representation of the dielectric permittivity: $\varepsilon^*(f,T) = \varepsilon' - i\varepsilon''$. The results were recorded isothermally for about 250 different frequencies at successive tested 193 temperatures. It made it possible to analyze the data in the representation $\varepsilon^*(f, T = const.)$ (Fig. 3), commonly used in *Critical Phenomena Physics* [47-49] and *Glass Transition Physics* [39, 50, 51, 61] and in the equivalent representation $\varepsilon^*(f = cont., T)$ (Fig. 4) often used in the *Physics of Ferroelectrics* [52] and *Relaxors* [20, 24].

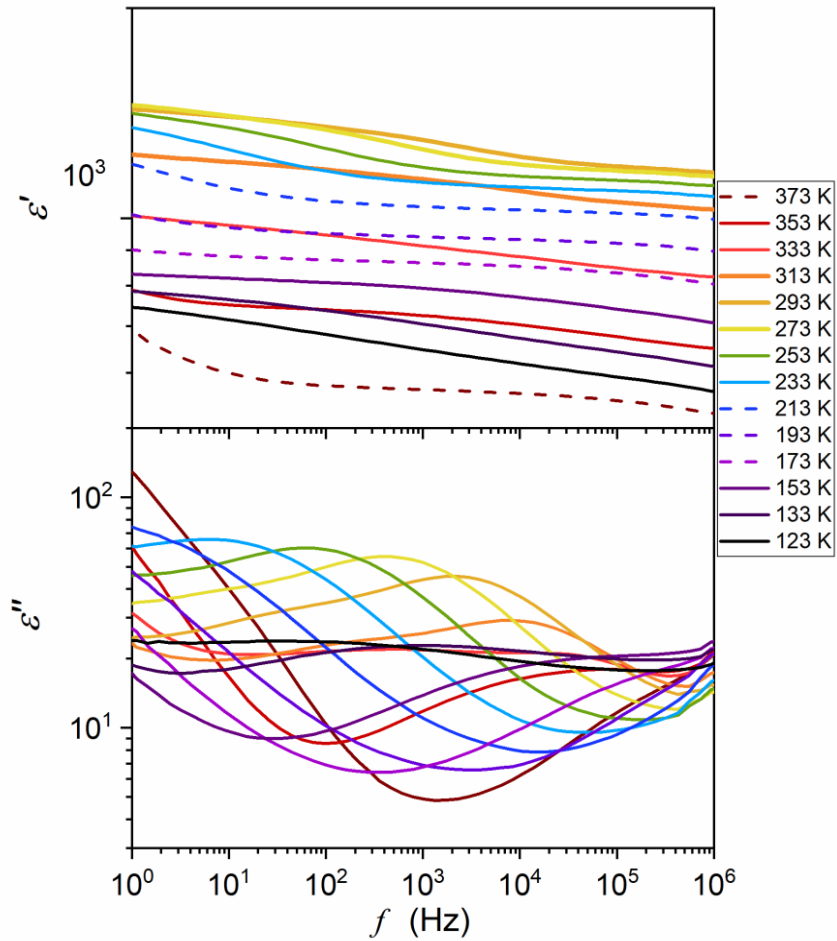

**Figures 3.** BDS related spectra showing frequency evolution for the real and imaginary part of dielectric permittivity (log-log scale), for selected temperatures in the tested ceramic specified in Table I. The complete set of data consists of 193 tested temperatures. See also the Appendix.

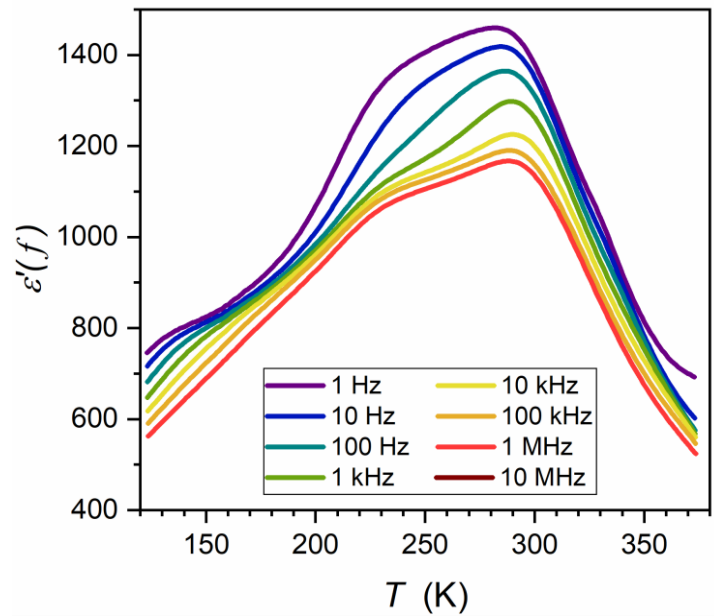

**Figure 4.** Temperature evolutions of the real part of dielectric permittivity for selected frequencies in the tested relaxor ceramic, specified in Table I. See also the Appendix.

## 3. Results and Discussion

**3.1a Model discussion: 'critical' view on dielectric constant related behavior in relaxor systems**

Curie-Weiss type scaling of dielectric constant temperature evolution is the essential experimental reference for basic 'homogeneous' ferroelectrics and related 'complex, heterogeneous' ferroelectric relaxor systems [5, 8, 9, 12, 13, 15, 19, 22, 23]. For interpreting CW type behavior, Devonshire [52-55] directly used the Landau model [56], which considers the free energy power expansion for the order parameter as the metric of appearing/disappearing element of symmetry on approaching the continuous phase transitions. Taking the electric polarization $P$ as the order parameter, one obtains [54]:

$$F = F_0 + \frac{a}{2}P^2 + \frac{b}{4}P^4 + \frac{c}{6}P^6 - EP \tag{11}$$

where coefficient $a = A(T - T_C)$; parameters $b$ and $c$ are considered approximately constant. The last term reflects the interaction with the electric field.

The above relation contains $\sim (c/6)P^6$ term, to include the tricritical point (TCP) case, the simplest multicritical point associated with meeting three critical points curves. For so-called symmetric TCP it manifests via the smooth crossover from discontinuous to continuous phase transition dependence [57, 58]. For the basic mean-field (MF) case, this term is absent. Eq. (11) yields the following pattern for pretransitional changes of the order parameter [47, 58]:

$$P(T) \propto (T_C - T)^\beta \tag{12}$$

The exponent $\beta = 1/2$ for MF and $\beta = 1/4$ for TCP

For the susceptibility, i.e., the order parameter changes by the coupled external field, $\chi = dP/dE$:

$$\chi(T) = \frac{a^{-1}}{(T-T_C)^\gamma}, \quad \text{for } T > T_C \tag{13a}$$

$$\chi(T) = \frac{(2a)^{-1}}{(T_C-T)^\gamma}, \quad \text{for } T < T_C \tag{13b}$$

The susceptibility-related exponent $\gamma = 1$, both for MF and TCP cases.

Notable that Eq. (11) leads to the prediction of heat capacity linear changes on both sides of $T_C$, i.e., no pretransitional anomaly associated with (critical) exponents and only the jump: $\Delta C_v = T_C a^2/2b$ [47] Such behavior does not correlate with experimental results, for which capacity pretransitional anomaly is evidenced [12, 13, 19]. The basic Landau-Devonshire model dependence (Eq. 11) [54], or generally the basic Landau model, which was exemplified for magnetization and paramagnetic-ferromagnetic transition [56], is related to the 'classic' behavior within the basic MF or TCP approximation, with a hypothetical negligible impact of pretransitional/precritical fluctuations. Notwithstanding, such an impact exists. To show it explicitly, Ginzburg supplemented the Landau equation with the gradient term [59, 60], directly recalling fluctuations. Implementing this concept to Eq. (11), one obtains:

$$F = F_0 + \frac{a}{2}P^2 + \frac{b}{4}P^4 + \frac{c}{6}P^6 + \kappa(\nabla P)^2 - EP \tag{14}$$

where $\kappa$ is the stiffness coefficient and the term $(\nabla P)^2 \propto \langle \delta P^2 \rangle$ is related to fluctuations of the order parameter around some 'equilibrium' value.

Eq. (12) yields temperature characterizations of the correlation length (size) $\xi$ and the lifetime $\tau_{fl.}$ of pretransitional/precritical fluctuations:

$$\xi(T) = \xi_0 |T - T_C|^{-\nu} \quad (15a) \qquad \tau_{fl.}(T) = \tau_0 |T - T_C|^{-\varphi} \propto [\xi(T)]^z \quad (15b)$$

where $\nu$ is the correlation length critical exponent, $\phi = z\nu$; z is the so-called dynamic exponent: $z = 2$ for the conserved order parameter and $z = 3$ for the non-conserved order parameter. For the classic behavior (MF, TCP): $\nu = 1/2$ and $\phi = 1$.

Eq. (12) leads to the following behavior of the heat capacity:

$$C_v(T \to T_C) \propto |T - T_C|^{-\alpha} \tag{16}$$

with exponents: $\alpha = 1/2$ ($T < T_C$) and $\alpha = 0$ ($T > T_C$) for MF; for TCP: $\alpha = 1/2$ both for $T < T_C$ and $T > T_C$.

Critical exponents are basic parameters characterizing pretransitional behavior. The grand success of the *Critical Phenomena Physics* was related to showing that their values depend only on the space ($d$) and the order parameter ($n$) dimensionalities [61]. Hence, microscopically different systems can be assembled into $(d, n)$ universality classes, in which isomorphic/equivalent physical properties are described by the same value of the exponents in the surrounding of critical (singular) points. This universal behavior splits into two categories: (***i***) non-classic, where exponents are small irrational numbers, and (***ii***) classic ones, where exponents are small integers or their ratios. The latter is associated with space dimensionalities $d \geq 4$ (single critical point, MF case) and $d \geq 3$ (the simplest multicritical point: TCP) [47, 58-61]. The 'classic' behavior is also linked to an 'infinite' range of intermolecular/inter-element interactions at the microscopic level. One can recall the Ginzburg criterion to comment on this issue and the interplay between classic and non-classic criticality. Implementing the discussion from refs. [59, 60] to the paraelectric – ferroelectric phase transition, one can link the classic behavior to the following form of the criterion:

$$\frac{\langle \Delta P^2 \rangle}{P^2} = \frac{1}{\xi^d} \frac{kT\chi}{P^2} < 1 \tag{17}$$

where $P$ has the meaning of the general order parameter and $\chi \propto |T - T_C|^{-\gamma}$ is for the order parameter coupled susceptibility.

The Ginzburg criterion shows that the classic – non-classic crossover can occur if the space-range associated with pretransitional fluctuations becomes smaller than the range of microscopic 'permanent' interactions (intermolecular, inter-element) characterizing a given system. It means that for systems with non-classic critical behavior, a crossover to the classic one may occur well remote from the critical point, where the correlation length drops enough. Indeed, such behavior was evidenced – for instance, a few tens of Kelvins away from the critical consolute temperature in binary critical mixtures of limited miscibility liquids ($d = 3, n = 1$ universality class: critical exponents $\gamma \approx 1.23, \beta \approx 0.325, \nu \approx 0.625$) [47, 62, 63]. However, in critical mixtures, the explicit classic behavior associated with exponents $\gamma = 1, \beta = 1/2, \nu \approx 1/2$, appeared in the broad surroundings of $T_C$ under the strong electric field or under the shear flow. Both agents cause precritical fluctuations' uniaxial elongation, which is possible in this domain under even moderate external impacts [64, 65]. In the given case, exogenous impacts do not affect intermolecular interactions, and the only factor leading to the 'anomalous' appearance of classic behavior may be local uniaxial symmetry in the given case induced by exogenic impacts. This concept led to the explanation of changes in the nonlinear dielectric effect (NDE), electro-optic Kerr effect (EKE) when approaching the critical consolute point and gas – liquid critical point. It also turned out to be crucial for explaining the mean-field nature of NDE, EKE, and dielectric constant pretransitional changes in the isotropic liquid phase of nematogenic liquid crystals, where rod-like uniaxial symmetry is the inherent feature. Recently, it was also used to show and explain NDE, EKE, and dielectric constant behavior in the liquid phase on approaching the orientationally disordered crystal (ODIC) phase of plastic crystals [64-67].

For explaining such behavior, essential is the inter-relation between meanings of the increased dimensionality ($d \geq 4$, for MF case) and the '~infinite' range of interaction. For both cases, it means that the number of nearest neighbors for a given molecule or element, which also means a possibility of interactions (''visibility'') is larger than the results from 'geometrical packing' of representing their spheres. Such a situation can occur if the local symmetry of elements responsible for the system or phenomenon is dominantly uniaxial. The above comment allows us to answer a fundamental question:

*"Why does the surrounding of the paraelectric-ferroelectric transition show the mean-field characterizations described by the Curie-Weiss 'law' (Eq.1), related to the exponent $\gamma = 1$ (Eq.13)?"*

In our opinion, it can be explained by the inherent uniaxiality of the ferroelectricity origin, associated with an uniaxial shift of charges within a basic element of the crystalline network.

As for the complex case of relaxor ferroelectric materials, one should take into account their basic material characterization, namely, they are built from micrometric size ($l_{Grain}$) grains, connected via 'molten' surfaces, can lead to partially amorphous inter-grain material. Consequently, one can assume that in the paraelectric phase of relaxor ceramics on cooling toward the para–ferro transition, first, the 'canonic' ferroelectricity develops within grains until the correlation length approaches the grain size. In the opinion of the authors, it can be associated with the Burns temperature $\xi(T_B) \sim l_G$. Further cooling towards the transition cannot increase the correlation length of pre-ferroelectric fluctuations up to the infinite terminal (Eq. 15a). However, they can improve the pre-ferroelectric ordering within limited volumes of grains. Consequently, one can expect the appearance of a strong local electric field. They can lead to some coupling of fluctuations restricted by borders of grains and can influence their interiors.

At this point, the temperature characteristics of the order parameter under the action of the coupled field, in the given case P and E, are worth recalling. For ferromagnetic systems, it is magnetization and magnetic field; for ferroelectric systems, it is electric polarization and electric field. With the permanent action of the external (global or local) field, the order parameter, instead of approaching zero to $T \to T_C$ according to Eq. (12) shows a strong deviation when passing from ferro- to para- phase, with a remaining non-zero value of the order parameter in the high-temperature para-phase. The onset and the value of this distortion depend on the field intensity. Following Eqs. (11) and (14) one obtains for dielectric constant and susceptibility:

$$\chi(T,P,E) = \varepsilon(T,P,E) - 1 = \left(\frac{\partial^2 F(T,P,E)}{\partial P^2}\right)^{-1} =$$

$$= \chi(T,P,E) - 1 = \frac{1}{a+3bP^2(E)} = \frac{1}{A(T-T_C)+3bP^2(E)} = \frac{A^{-1}}{T-(T_C+3A^{-1}bP^2(E))} \quad (18)$$

The local electric field arising from the ferroelectric arrangement within grains is not uniform in values and directions. Following Eq. (18), one can expect 'pseudospinodal temperatures' [68] matched with different maximal available dielectric permittivity values.

The authors stress that similar functional forms of pretransitional behavior are related to Eqs. (1), (13), and (18) for $T \to T_C$, Eqs. (1) and (14)), and the pseudospinodal temperature: $T \to T_{SP} = T_C + (3A^{-1}bP^2(E))$ (Eq. (18)). However, the latter is associated with finite terminal dielectric permittivity / dielectric constant values.

### 3.1b Model discussion: 'critical' view on dynamics in relaxor systems

'Glassy' dynamics is the next unique feature of relaxor ceramics [8, 9, 13, 18, 19, 22-25]. It is proved by portraying the evolution of the primary relaxation time by VFT relation (Eqs. 3 and 4), instead of the simple Arrhenius pattern $\tau(T) = \tau_\infty exp(E_a/RT)$ with $E_a = const$, and the non-Debye changes of loss curve shape, for which the HN (Eq. 5) is recalled. Such scaling recalls the pattern occurring in the previtreous domain (i.e., above the glass temperature $T_g$) of glass-forming liquids. The origin for these universal changes related to $\tau(T_g) < 100s$ time scales remain a challenge [39, 50]. For relaxor systems, they are explicitly related to approaching the paraelectric-ferroelectric transition. They can be associated with the development of pretransitional fluctuations time-scale (Eq. 15b), which can be paralleled by a single dipole moment relaxation due to the MF nature of the phenomenon. Below $T_B$, which we associate with the reaching by the correlation length (Eq. 15a), the size of the grain ($l_G$) increasing frustration associated with this fact and growing internal local electric fields can appear. Interestingly, passing $T_B$ temperature seems not to affect the parameterization of $\tau(T)$ using the VFT relation.

Recently, however, it has been shown that the VFT relation is primarily important as an effective description tool for glass-forming systems [39].
The insight based on the analysis of the apparent activation energy index $I_L(T) = -dlnE_a(T)/dlnT$ led to the following expression for changes in configurational entropy [39, 69]:

$$S_C(T) = S_0 t^n = S_0 \left(\frac{T-T_K}{T}\right)^n = S_0 \left(1 - \frac{T_K}{T}\right)^n \quad (19a)$$

$$lnS_C(T) = lnS_0 + nlnt \Rightarrow \left(\frac{dlnS_C(T)}{d(1/T)}\right)^{-1} = \left(\frac{1}{nT_K}\right) + n^{-1}T^{-1} \quad (19b)$$

where $S_0 = const$, $T_K$ is related to the so-called Kauzmann temperature, the exponent $0.18 < n < 1.6$; the upper limit is related to the dominance of the orientational order and the lower one - the translational order. The case $n = 1$ is for systems with no preferable type of symmetry.

It leads to the following 'VFT-extended' equation [39, 69-71]:

$$\tau(T) = \tau_\infty exp\left(\frac{D}{T}t^{-n}\right) = \tau_\infty exp\left(\frac{DT^{n-1}}{(T-T_K)^n}\right) \tag{20}$$

It correlates with the VFT equation for $n = 1$, but the analysis of experimental data showed that for relaxor systems $n > 1$. However, the general Eq. (20) contains four fit parameters, which significantly burdens the reliability of the analysis. A solution may be to define the $n$ parameter independently, for example, using the configurational entropy analysis, as defined by Eq. (19b). Determining changes in structural entropy requires exact and long-range experimental heat capacity results, which are hardly available.

Recently, however, it has been shown that a universalistic description of the so-called steepness index. $m_T(T) = dlog_{10}\tau(T)/d(T_g/T)$, which is proportional to the apparent activation enthalpy $H_a(T) = dln\tau(T)/d(1/T)$ [72]:

$$m_T(T) = \frac{dlog_{10}\tau(T)}{d(T_g/T)} = \frac{1}{T_g ln10}\frac{dln\tau(T)}{d(1/T)} = C \times H_a(T) = C\frac{M}{T-T_g^*} \tag{21}$$

where $C, M = const$ and $T_g^* < T_g$ is the extrapolated singular temperature.

The above relation directly leads to the following three-parameter relation [72]:

$$\tau(T) = C_\Gamma\left(t^{-1}exp(t)\right)^\Gamma \tag{22a}$$

$$ln\tau(T) = lnC_\Gamma + \Gamma(t - lnt) \tag{22b}$$

where $t = (T - T_g^*)/T$ and $C_\Gamma = const$.

Notably, the number of adjustable parameters can be reduced to only two, since $T_g^*$ can be determined via scaling Eq. (19), using the linear regression for experimental data presented in the plot $H_a^{-1} = (dln\tau(T)/d(1/T))^{-1}$ vs $T$. Knowing $T_g^*$ one can present experimental data using the plot defined by Eq. (22b), namely $ln\tau(T)$ vs. $t - lnt$, and using the linear regression fit, determine $C_\Gamma$ and $\Gamma$ parameters. Hence, for portraying $\tau(T)$ via Eq. (22a) the nonlinear fitting can be avoided, and the reliable estimation of optima values of parameters, including their errors, is possible.

Eq. (22) links features of the 'activated' (i.e., SA-type: Eq. (3a)) and the critical-like behavior. Notable is the link of the exponent $\Gamma$ to the dominated local symmetry in the given system. If the uniaxial or translational symmetries are dominant, Eq. (22a) can be fairly approximated by the critical-like relation [72-78]:

$$\tau(T) = \tau_0(T - T_C^*)^\varphi \tag{23}$$

where the exponent $\varphi \approx 9$ and $T_C^* < T_g$

It is notable that Eq. (23) correlates with the so-called dynamical scaling model (DSM) [79] check-point equation with the exponent $\varphi = 9$, suggested as 'universal', at least for glass forming low molecular weight liquids and polymers. Such a statement has not found reliable experimental confirmation. However, the authors of this work (ADR, SJR) showed, using distortions-sensitive analysis, that Eq. (23) perfectly describes liquid crystalline (LC) systems, with a clear uniaxial symmetry of molecules.

We emphasize this fact because DSM is an inherently mean-field model, which is also the feature of the mentioned LC systems, coupled with the uniaxility. [39, 72-78]
The discussion presented in this section indicates that the VFT relation used standardly for describing 'glassy dynamics' in relaxor systems, is a tool for an effective description, and inferences based on it may have limited fundamental significance. We note the role of the critical-like, mean-field description and the importance of uniaxial symmetry, which correlates with the discussion on static properties, dielectric susceptibility & dielectric constant in section 3.1a.

**3.1c Model discussion: 'critical' view on Clausius-Mossotti local field in ferroelectric systems**

Shortly after Michel Faraday introduced the dielectric constant for characterizing the properties of dielectrics, this quantity became particularly important for obtaining fundamental insight into the microscopic properties of such materials [80]. In 1850, Mossotti introduced the first local field concept, which, after additions introduced by Clausius, is now known as the local field Clausius-Mossotti model [81-83]. Referring to further development concepts in this direction, a molecule/element inside a dielectric subjected to an external electric field $E$, for example, by placing it in a capacitor, is affected by an effective local field [82]:

$$F = E + E_1 + E_2 \tag{24}$$

where $E_2$ is for the electric field created by elements/molecules within a semi-microscopic cavity surrounding a given molecule/element, and $E_1$ results from charges situated on the surface of the cavity.

For dielectric (gas or liquid with a random) distribution of elements or a regular crystalline lattice: $E_2 = 0$. Summarizing the impact of the cavity surface charge yields, one obtains [82]:

$$E_1 = P/3\varepsilon_0 \tag{25}$$

where $P$ denotes the polarization vector and $\varepsilon_0 = 8.854\ (pFm^{-1})$ is the vacuum electric permittivity.

Such approximation can be applied for gas dielectrics with non-interacting molecules or non-dipolar liquids [81, 82]. Recalling the dielectric displacement vector: $D = \varepsilon_0 E + P = (\chi' + 1)\varepsilon_0 E = \varepsilon_0 \varepsilon' E$ and the link between the polarizability vector and the basic element/molecule polarizability: $P = \varepsilon_0 \chi' E = N\alpha_p F,$ with $\alpha$ meaning the basic element /molecule polarizability and $N = \rho N_A M^{-1}$ is for the number of basic elements/molecules per unit volume ρ denotes density, $M$ the molecular mass, and $N_A$ is for the Avogadro number, one obtains [82]:

$$F = \frac{P}{3\varepsilon_0} = \frac{\chi'}{\chi'+3} P \quad \Rightarrow \quad \chi' = \frac{N\alpha_P}{3\varepsilon_0} \frac{\chi'}{\chi'+3} \tag{26}$$

The re-arrangement of the latter yields:

$$\chi' = \frac{P}{\varepsilon_o E} = \frac{N\alpha_P/\varepsilon_0}{1-N\alpha_P/3\varepsilon_0} \tag{27}$$

The above discussion (Eqs. 24-27) are canonical results presented in classic monographs on dielectric physics. Von Hippel supplemented them by considering the relation (27) in dipole dielectrics, especially liquid ones, using the relation introduced by Debye $\alpha_P = \mu^2/3k_B T$, which transformed Eq. (27) to the form [81, 82]:

$$\chi = \varepsilon - 1 = \frac{3T_C}{1-T_C} \qquad (28)$$

where $T_C = N\mu^2/9k_B\varepsilon_0$.

Von Hippel, in his classic monograph, pointed out the paradoxical consequences of this reasoning for such common dipolar dielectric as water, showing that it leads to the paraelectric – ferroelectric transition $T_C \approx 1520K$, concluding [81]: *'water should solidify by spontaneous polarization at high temperature, making life impossible on this earth*!'. This paradox result is often cited in monographs and course lectures for students because of its exceptional impressiveness, showing the consequences of exceeding the basic assumptions of a given model.

Von Hippel associated it with the need to take into account short-range interactions, which he associated with non-zero field *E₂*, and suggested the switch to Onsager-related model approaches, reducing the cavity to a size similar to that of a molecule. The paradox anomaly for dielectric liquids has been removed by taking short interactions, and for interpreting experimental data, the Kirkwood or Froelich models are used [82].

It is worth mentioning here, however, that the example of von Hippel's paradox ignores an important fact. It assumes the density of water for 'normal' conditions, i.e., *d* = 1 *g/cm³* [81, 82]. For such a density to exist in 'paradoxical conditions', it would be necessary to enclose a given volume of water in a pressure capsule and heat it above 1500 K, which has to create multi-GPa pressure. It can yield even exotic properties, which are often obtained for materials under extreme pressures.

The following summary from the monograph *Dielectric Physics* by Chełkowski can summarize the considerations regarding the application of the Clausius-Mossotti local field model [82]: *'(…) it is obvious that in the case if dipolar materials (…) the Lorentz field model cannot be employed*'.
However, the Clausius-Mossotti model is a widely accepted fundamental concept describing the properties of ferroelectric materials, the primary experimental confirmation of which are changes in the dielectric constant described by the Curie-Weiss relation. Notwithstanding, there are materials in the solid phase (classic ferroelectrics) or liquid phase (liquid crystalline ferroelectrics) inherently associated with significant dipole moments where Mossotti Catastrophe, related to Curie-Weiss Eq. (1), is the basic property [36-38, 52]. Several models are addressing this problem, essentially referring to the qualitative explanation of von Hippel [81], who stated that in ferroelectric materials, an applied electric field or thermal motion can yield a charge displacement and, consequently, a net dipole moment within the crystalline network, which can be further increased due to supplementary displacement caused by inter-ions couplings. The process continues until the thermal agitation can be overcome at a critical temperature, and the Mossotti Catastrophe, paralleling the Curie-Weiss relation, can occur.

However, there is still a need for a simple answer to the simplest question of why this situation occurs only in ferroelectric materials (solid or liquid) and not in classic dielectric liquids. Heuristically, the answer to this question can be that short-range interactions (omitted in the Clausius-Mossotti model) are important, as discussed above.

The '*Critical*' discussion presented in sections 3.1a and 3.1b provides a simple answer. The intrinsic bond of basic ferroelectrics with uniaxial symmetry leads to the appearance of mean field properties. It means that it is a kind of 'immersion' of the induced moments in the mean-field that characterizes the effective interactions so that special microscopic features that may appear in the interaction of neighboring moments disappear. As a result, a kind of 'effective gas' of independent dipole moments is created, which correlates with the basic assumptions of the Clausius-Mossotti local field. Significant distortions from this picture associated with the specific material characterization appear in the broad surrounding of the paraelectric – ferroelectric transition for relaxor ceramics.

## 3.1 Experimental results and discussion

Studies were carried out in $Ba_{0.65}Sr_{0.35}TiO_3$ relaxor ceramic (99.3%), whose preparation and characterization are described in the Experimental section. It also contains master plots, showing frequency-related ($T = const$: Fig. 3) and temperature and temperature-related ($f = const$: Fig. 4) master plots for the real and imaginary components of dielectric permittivity. They have been selected from data covering 193 tested temperatures in the range $123\,K < T < 373\,K$, to illustrate general features. Dielectric constant is the basic property for which temperature evolution is considered for relaxor ceramics. It is defined as the near-constant value of $\varepsilon' = \varepsilon$ in the static frequency domain where a frequency shift has a negligible impact on detected values. It is visualized as the horizontal domain at $\varepsilon'(f, T = const)$ spectrum, for dipolar dielectrics usually for $1kHz < f < 10MHz$ [51].

Figure 3 shows that such behavior is almost absent for the tested relaxor ceramic, particularly near the paraelectric – ferroelectric transition. It is indicated by 'thicker' curves in Fig. 3. The static-type horizontal behavior appears only well above the transition (for the isotherm $T = 373K$) and for $T \approx 200K \pm 30$. Notable that the Curie-Weiss temperature $T_C \approx 292K$.

Consequently, the discussion of the Curie-Weiss behavior for relaxor ceramics should be carried out in frames of the real part of dielectric permittivity, and 'dielectric constant' should be treated as the replacement name. For such meaning of 'dielectric constant', the frequency $f = 10kHz$ can be a reasonable selection, often used in studies on relaxor systems.

Figures 5 – 7 present the behavior of $\varepsilon'(T, f = 10kHz)$, focused on testing the temperature evolution with the support of the distortions-sensitive and derivative-based analysis [39, 42-44, 64, 65]. It has already been used in glass-forming systems and 'critical' liquids, revealing significant features hidden for the direct nonlinear fitting of experimental data.

Figure 5 shows the temperature change of 'dielectric constant in the temperature range covering 200 K, including the evolution of its reciprocal. It recalls the commonly applied analysis, recalling

Curie-Weiss Eq. (1), and also used for determining the Burns temperature $T_B$ linked to the distortion from CW behavior in the route to paraelectric – ferroelectric transition. The departure from CW Eq. (1) occurs gradually, and precise estimation of its value is not possible, namely: $T_B = 340K \pm 5K$. Notable are linear changes of $1/\varepsilon(T)$ in the paraelectric phase, which can be considered as confirmation of the process description via Curie-Weiss Eq. (1). It extends for ca. 50K, although a weak distortion on approaching the high-temperature terminal $(T \approx 375K)$ seems to emerge.

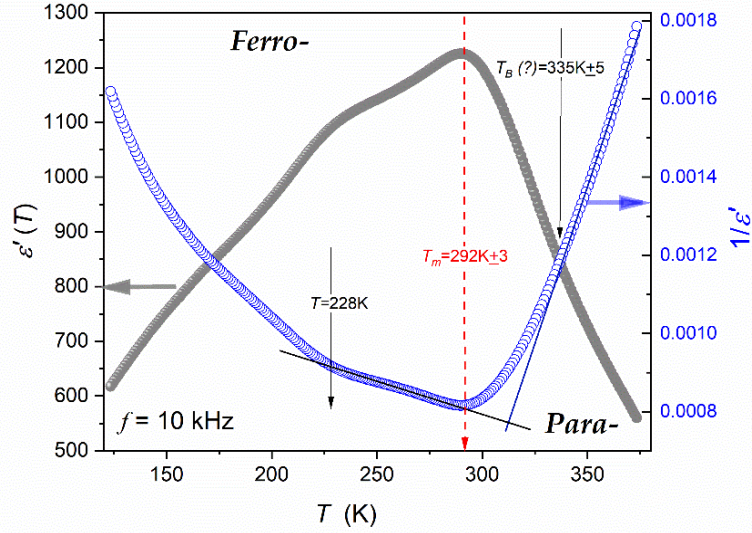

**Figure 5.** Temperature changes of the real part of dielectric permittivity, related to 'dielectric constant', also shown for its reciprocal. Results for $Ba_{0.65}Sr_{0.35}TiO_3$ relaxor ceramic.

The precise determination of the $T_B$ value and ultimate validation of the CW description can be expected using distortions-sensitive data analysis using Eq. (1b):

$$\frac{d(1/\varepsilon(T))}{dT} = \frac{d(A_{CW}^{-1}T - A_{CW}^{-1}T_C)}{dT} = A_{CW}^{-1} = const \qquad (29)$$

Such analysis is presented in Figure 6: the horizontal line expected according to Eq. (29), occurs only for the ferroelectric side of the curve related to the paraelectric – ferroelectric transition. There is no horizontal line for the paraelectric side, which is the focus of studies in relaxor systems: the validation of CW description is negative (!)

Notably that in the ferroelectric phase, near $T \approx 170K$, a hallmark of the next phase transition appears. For $T > 170K$ it follows the pattern parallel to Eq. (29), for ca. 40K.

Figure 7 presents the semi-log scale for experimental data from Fig. 5, supplemented by the distortions-sensitive and derivative-based analysis. It has two targets. The first is the validations of (surprising) fair exponential behavior expending from $T = 375K$ to $T = 315K$:

$$\varepsilon(T) = \varepsilon_{ref.}exp(a'T) \quad \Rightarrow \quad ln\varepsilon(T) = ln\varepsilon_{ref.} + a'T \qquad (30)$$

where $\varepsilon_{ref.}, a' = const.$

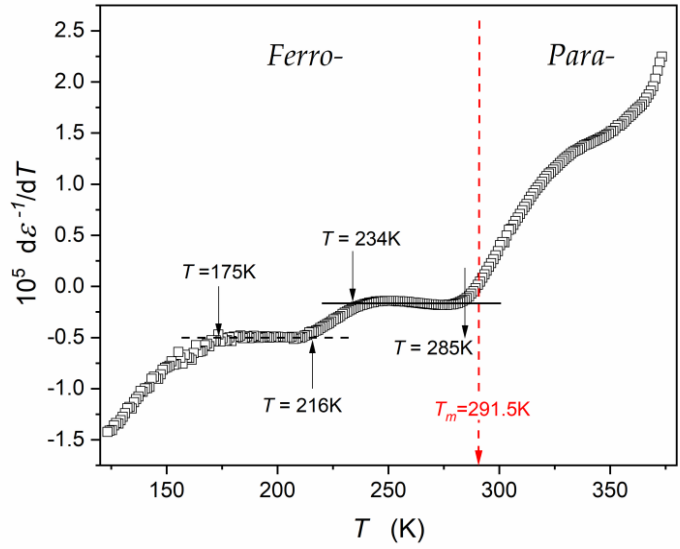

**Figure 6.**   Temperature changes of the 'dielectric constant' reciprocal derivative, focused on the distortions-sensitive test of the Curie-Weiss behavior, manifesting via horizontal lines. The analysis for $Ba_{0.65}Sr_{0.35}TiO_3$ relaxor ceramic - based on experimental data shown in Fig. 5.

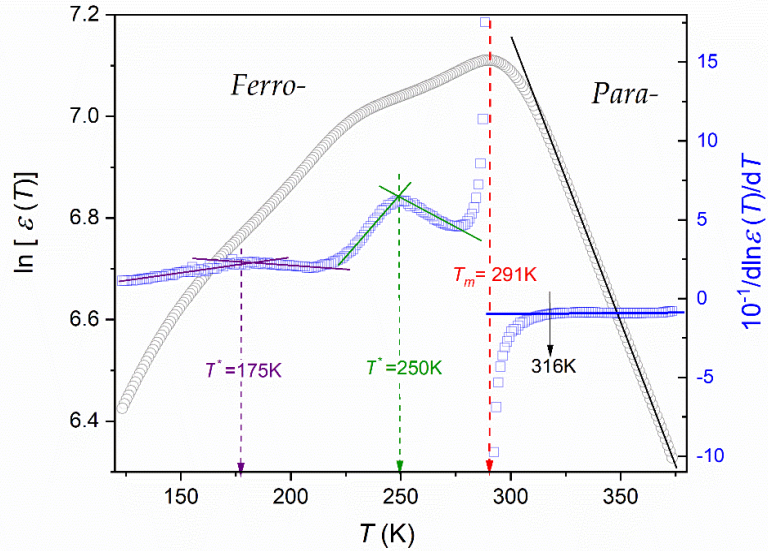

**Figure 7.**   Temperature changes of the 'dielectric constant' logarithm and the reciprocal of its derivative for the distortions-sensitive test of such behavior, which is manifested by the horizontal line. The analysis for $Ba_{0.65}Sr_{0.35}TiO_3$ relaxor ceramic, based on experimental data from Fig. 5.

It is confirmed by the solid line following experimental data in the paraelectric phase. It is supplemented by the distortions-sensitive and derivative-based analysis, presented as $[dln\varepsilon(T)/dT]^{-1}$ vs. $T$ analytic plot. It enables the 'subtle' test of the existence of critical-like domains, described as follows:

$$\varepsilon(T) = \varepsilon^0|T - T^*|^{-\phi} \; \Rightarrow \; ln\varepsilon(T) = ln\varepsilon^0 - \phi ln|T - T^*| \; \Rightarrow \; \frac{d(ln\varepsilon(T))}{dT} = \frac{-\phi}{|T - T^*|} \; \Rightarrow$$

$$\left[\frac{d(ln\varepsilon(T))}{dT}\right]^{-1} = -\phi T \mp \phi T^* = a + bT \qquad (31)$$

where $\varepsilon^0, a, b = const$, $T^*$ is for the critical-like temperature, and $\phi$ is the 'critical' exponent.

The mentioned plot also enables the validation of Eq. (30), yielding a horizontal line, namely:

$$\left[\frac{d(ln\varepsilon(T))}{dT}\right]^{-1} = (a')^{-1} = const \qquad (32)$$

Such also enables a precise estimation of singular temperatures related to phase transition.

The interesting feature is the agreement of Eq. (30) with the output model-relations proposed in ref. [6]. Figure presents results of the derivative-based analysis of 'dielectric constant; changes in the surrounding of its maximum, related to the transition from the paraelectric to the ferroelectric phase. The linear domain detected in such analysis is related to (Fig. 8):

$$\frac{d(ln\varepsilon(T))}{dT} = a + bT \;\; \Rightarrow \;\; d(ln\varepsilon(T)) = (a + bT)dT \qquad (33)$$

The integration of the above yields:

$$\varepsilon(T) = A\exp(c + aT + bT^2) \qquad \text{for} \qquad 285K < T < 314K \qquad (34)$$

i.e., for the surround of paraelectric – ferroelectric transition

Notable that for the paraelectric side of the transition, the following portrayal was validated (Fig. 7):

$$\varepsilon(T) = A\exp(b + aT) \qquad \text{for} \qquad 315K < T < 375K \qquad (35)$$

For the ferroelectric side of the transition:

$$\varepsilon(T) = \frac{C}{|T-T_C|} \qquad \text{for} \qquad 234K < T < 285K \qquad (36)$$

i.e., correlated with the mean-field Landau-Devonshire model [54, 56].

Notably, there are almost no 'gaps' between descriptions related to subsequent temperature domains. Temperature changes of the imaginary part of dielectric permittivity for the discussed 'quasi-static' frequency $f = 10kHz$ is shown in Figure 9. This magnitude reflects the energy absorbed for subsequent processes, supplementing the message from the scan of the real component, reflecting mainly the appearance and arrangement of permanent dipole moments. For the evolution of $\varepsilon''(T)$ it appears in the paraelectric phase, for $T = 330K - 345K$, but it is entirely invisible $\varepsilon'(T)$, i.e., the 'dielectric constant'. Also, in the ferroelectric phase, there is a strong manifestation of relaxation processes, which for $\varepsilon'(T)$ they become explicit only for disturbances-sensitive & derivative-based analysis. The mentioned evidence is even stronger, especially in the paraelectric phase for $tan\delta = \varepsilon''/\varepsilon'$, which can be related to the fact that this quantity can also be defined as $D = tan\delta = energy\ lost\ per\ cycle/energy\ stored\ per\ cycle$ (in Fig. 9 the cycle is related to $f = 10kHz$), i.e., it determines the energy of the process itself, minimizing the influence of the 'background', i.e., of the entire system [80, 81]. This property is also called the dissipation factor, used for defining the quality factor $Q = 1/D$, used in applications.

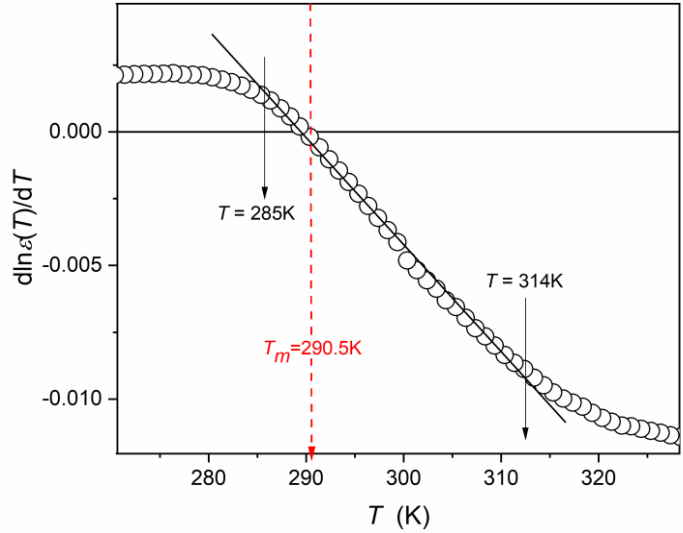

**Figure 8.** Temperature changes of the derivative of 'dielectric constant' ( $\varepsilon'(f = 10kHz)$ ) logarithm in the surroundings of the paraelectric–ferroelectric transition. The dashed red line indicates the temperature of the 'dielectric constant' maximal value. Solid, black arrows indicate terminals of the linear behavior. The analysis for $Ba_{0.65}Sr_{0.35}TiO_3$ relaxor ceramic - based on experimental data shown in Fig. 5.

Figure 9 shows that the tested system is characterized by a relatively low dissipation/loss factor. It increases on approaching the paraelectric – ferroelectric transition, which can be associated with an increasing number of permanent dipole moments able to interact with the external electric field, also coupling within multi-element fluctuations, which is associated with anomalously increasing susceptibility $\chi = \varepsilon - 1$ reflecting the rising sensitivity of local order parameter changes (polarizability) to the electric field. Such impact diminishes away from the transition. The impact of changing frequency in temperature scans for the tested temperature range is presented in Figure 10.

In such a way, the significant uncertainty for their determination via the Havriliak – Negami relation [51], requiring the nonlinear fitting, is avoided. Such fitting is associated with at least four adjustable parameters, and their number increases to eight if the merge of two relaxation processes creates the loss curve. Loss curves for characteristic temperature domains, with indications of basic relaxation processes and coupled relaxation times, are shown in Figure 11.

Relaxation times that appear in dielectric permittivity spectra were determined from peak frequencies of loss curves $\tau = 1/2\pi f_{peak}$, supported by the analysis of $dlog_{10}\varepsilon''(T)/dT$ and $dlog_{10}\varepsilon''(f)/dlog_{10}f$ enabling its unequivocal estimation.

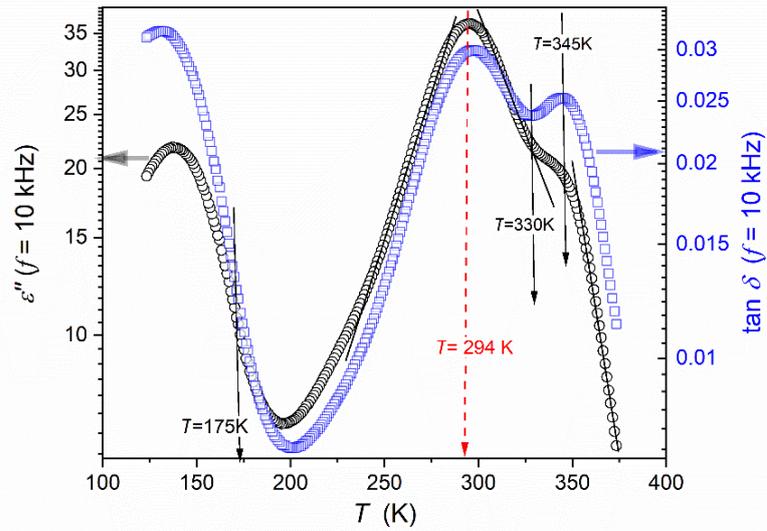

**Figure 9.** Temperature changes of the imaginary part of dielectric permittivity ($\varepsilon''(f = 10 kHz)$ ) and related $\tan \delta = \varepsilon''/\varepsilon'$. Solid, black line indicates characteristic temperatures, and the dashed red line is related to the paraelectric- ferroelectric transition: note a slight shift in comparison with temperatures detected in $\varepsilon'(T)$ analysis. The results are for $Ba_{0.65}Sr_{0.35}TiO_3$ relaxor ceramic – (see Fig. 3 and the Appendix).

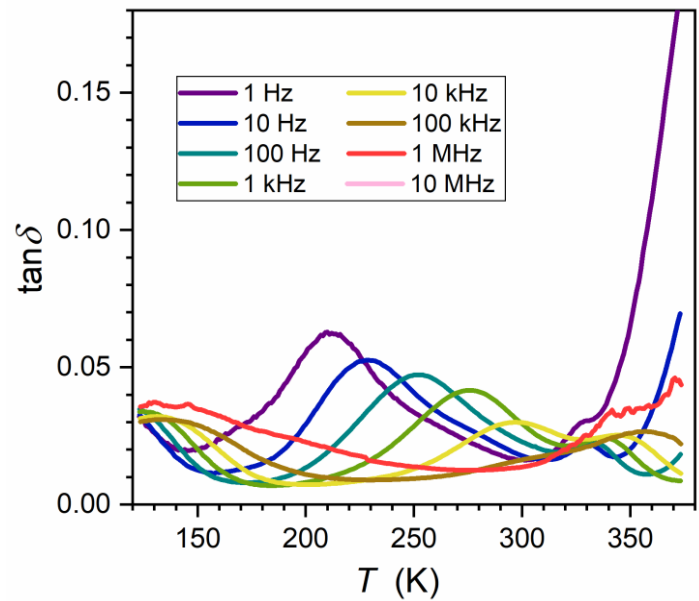

**Figure 10.** Temperature evolutions of $tg\delta(T, f = const) = \varepsilon''(f,T)/\varepsilon'(f,T)$ for selected frequencies - in the tested relaxor ceramic, specified in Table I.

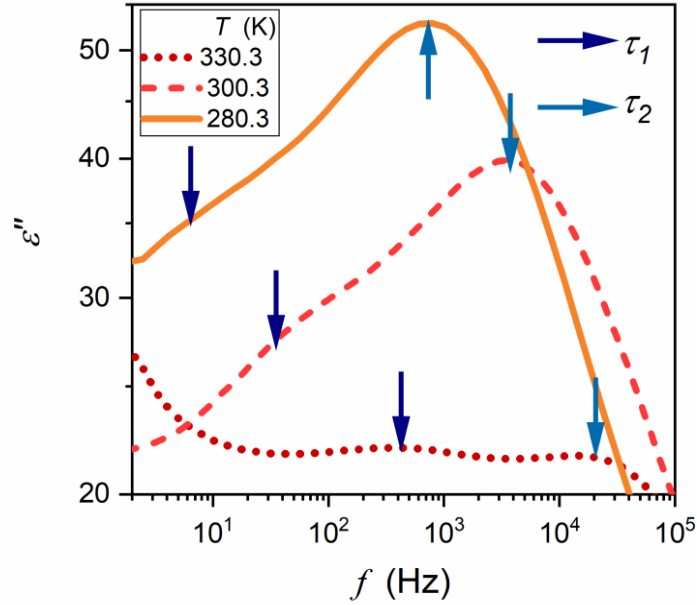

**Figure 11.** Dielectric loss curves in the paraelectric phase for three selected temperatures. Relevant relaxation processes are indicated. Results are for $Ba_{0.65}Sr_{0.35}TiO_3$ relaxor ceramic.

Figure 12 presents the obtained map of relaxation times using the Arrhenius scale $log_{10}\tau(T)$ vs. $1/T$ The inset is for the relaxation time at low temperatures in the ferroelectric state. It appears that the tested system exhibits a unique feature. Usually, the Super-Arrhenius behavior occurs in the paraelectric phase and terminates in the vicinity of $T_m$. For the tested compound it terminates at $T_{term.} \approx 330K$. Notable is the Super-Arrhenius behavior, related to Eq. (19), which has been shown via the apparent activation enthalpy tests focused on validating its portrayal by Eq. (18). This result is presented in Figure 13.

On further cooling towards the transition, a new process emerges. It follows explicitly the simple Arrhenius pattern, with the constant activation energy extending deeply into the ferroelectric state, without a hallmark when passing $T_m$ temperature (Figure 12). The height (maximum) of related loss curves strongly increases on cooling, as presented in Figure 3 and Figure 14. Figure 15 presents the scaled superposition of $\tau_2$ relaxation time-related loss curves, showing the essentially non–Debye and broad distribution of relaxation times.

We would like the detection of phase transformations in the ferroelectric phase, visible for temperature evolutions of 'dielectric constant' (Fig. 7) , which suggests a link to the arrangement of permanent dipole moments and also for $\varepsilon''(T)$ and $tan\delta(T)$ which can reflect the energy loss associated with these phenomena. The process related to the lowest temperature introduces the additional relaxation time, shown in the inset in Figure 12, and follows the basic Arrhenius pattern for the temperature evolution.

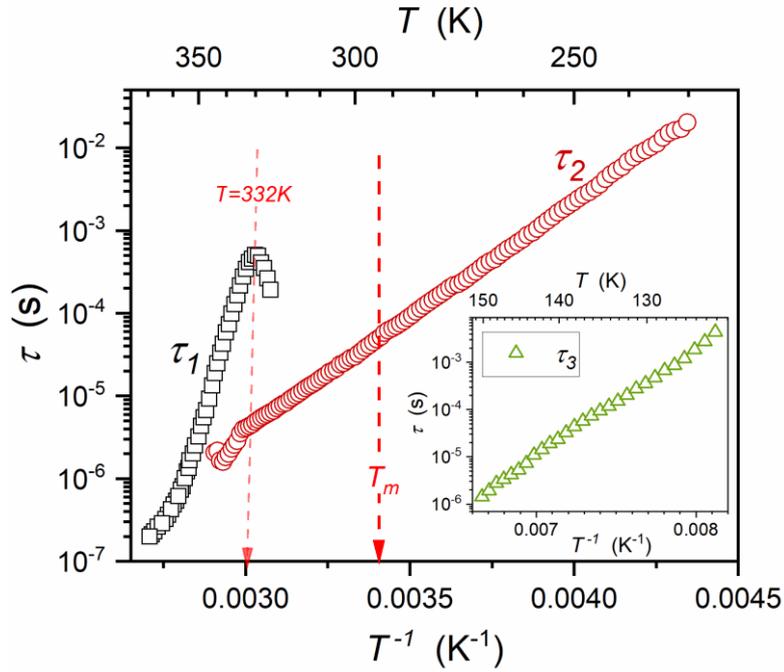

**Figure 12.** Arrhenius plot for relaxation times detected in $Ba_{0.65}Sr_{0.35}TiO_3$ relaxor ceramics. The inset changes of the relaxation for the process emerging in ferroelectric phase at low temperatures.

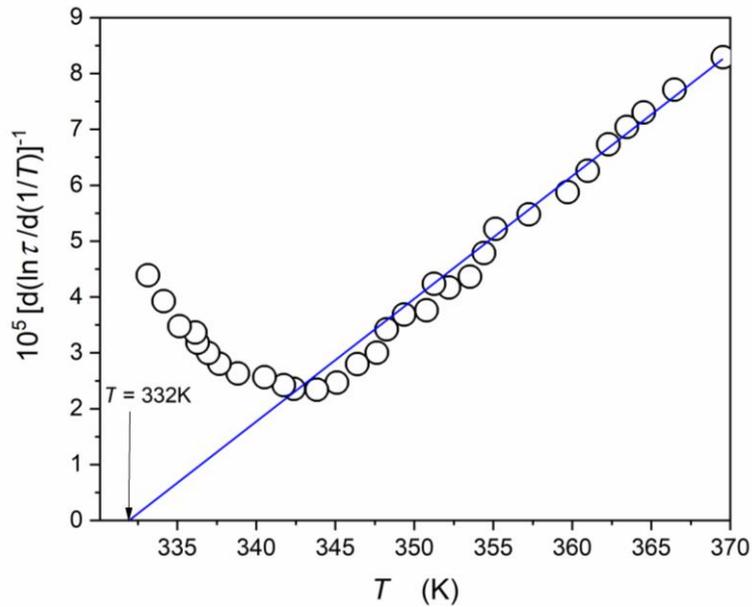

**Figure 13.** The temperature dependence of the reciprocal of the apparent activation enthalpy focused on validating Eqs. (21) and (22), which should manifest as the linear behavior. The singular temperature $T^*$ is indicated by the arrow.

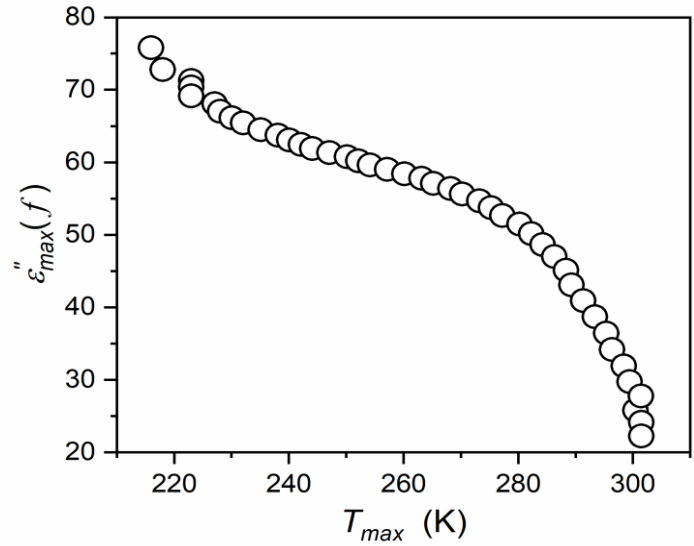

**Figure 14.** Temperature changes of the maxima of $\tau_2$ - relaxation time- related loss curves, as indicated in Figure 12.

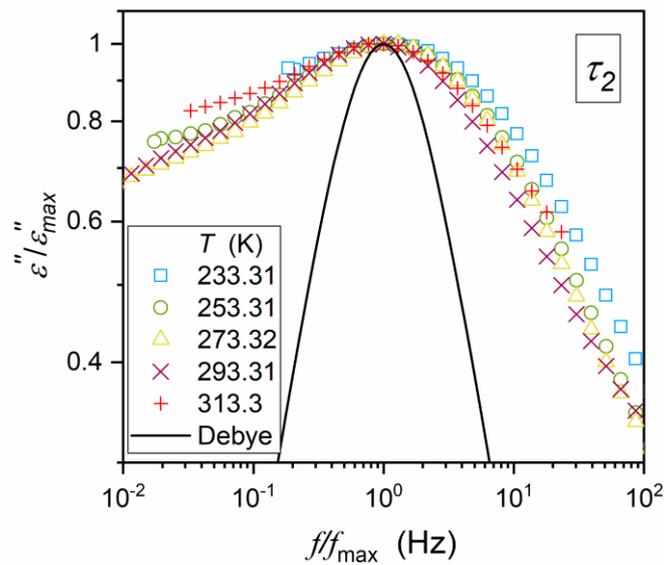

**Figure 15.** Time-temperature-superposition (TTS) of relaxation time in the tested relaxor ceramic, covering both paraelectric and ferroelectric phases. For comparison, the single relaxation time-related Debye distribution is also shown. The plot is presented in the log-log scale.

For applications of relaxor systems, the sensitivity of dielectric properties, particularly 'dielectric constant', to the external electric field is essential. Fundamental origins of such behavior also have remained a challenge. Figure 16 shows such behavior for the relaxor ceramic discussed in the given report. Figure 17 presents the same experimental data to show relative changes of 'dielectric constant' in respect to the no-field case ($U = 0, E = 0$). Notably, relatively strong changes of 'dielectric constant' occur for relatively weak electric fields.

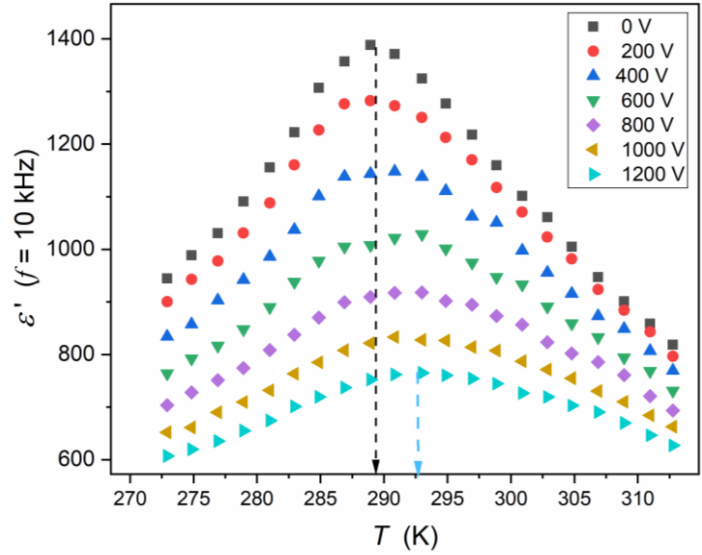

**Figure 16.** Changes of 'dielectric constant' ($f = 10\ kHz$) for $Ba_{0.65}Sr_{0.35}TiO_3$ sample, specified in Table I, using for a disk with $h = 1mm$ height and voltages given in the figure. The arrows indicate maximal values.

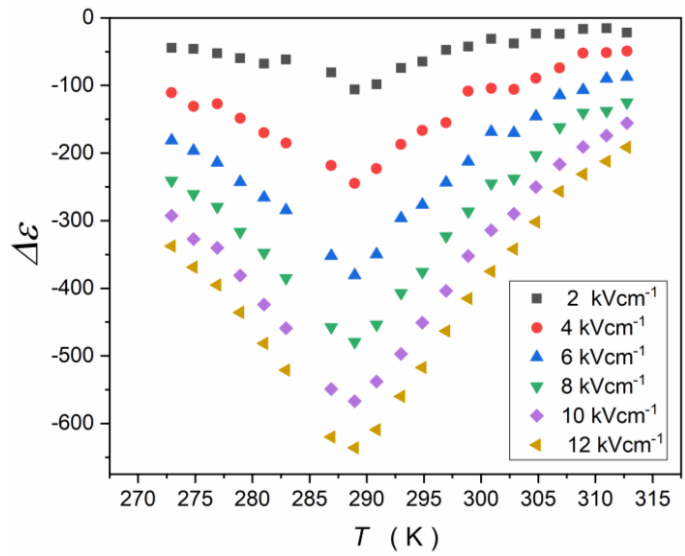

**Figure 17.** Relative changes of 'dielectric constant' ($f = 10\ kHz$) for $Ba_{0.65}Sr_{0.35}TiO_3$ sample (specified in Table I), comparing scans under electric field $E \neq 0$ to $E = 0$ behavior.

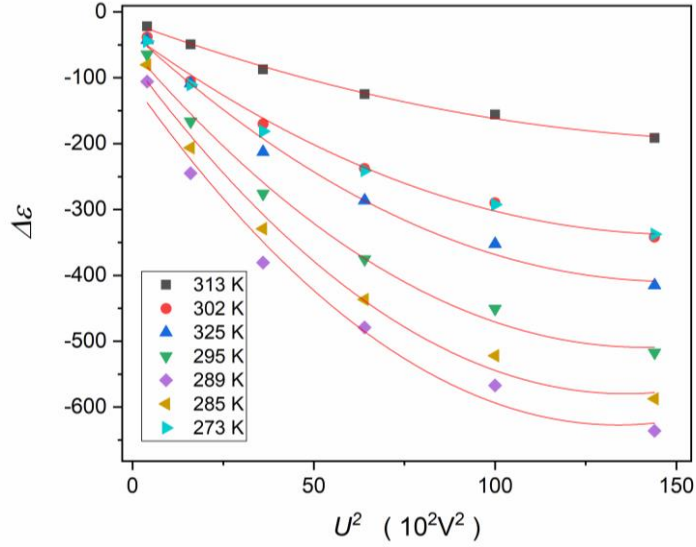

**Figure 18.** Relative changes of 'dielectric constant' ($f = 10\ kHz$) in Ba$_{0.65}$Sr$_{0.35}$TiO$_3$ sample (specified in Table I), versus the square of the applied voltage to $h = 1mm$ 'thick' sample.

Worth indicating is also a relatively large shift if $\varepsilon(T)$ curve maximum reaching $\Delta T(E) \approx 3K$, for $E = 12 kV cm^{-1}$, which shows the ability for the electrocaloric effect in the given system.

Figure 18 presents the test of the electric field intensity, or alternatively the applied voltage, of $\Delta\varepsilon(E) = \varepsilon(E = 0) - \varepsilon(E)$, in the surroundings of the paraelectric–ferroelectric transition. Red curves show that the following polynomial can portray experimental data:

$$\Delta\varepsilon(E) = \varepsilon_{ref.} + aE^2 + bE^4 \qquad (37)$$

This report shows that it is possible to describe the temperature changes in the dielectric constant: (*i*) in the ferroelectric phase (Eq. 36), (*ii*) in the environment of the diffused, stretched in temperature, paraelectric – ferroelectric transition (Eq. 34), (*iii*) in the paraelectric phase (Eq. 35). The transition to subsequent domains when the temperature changes takes place without a significant temperature gap. It allows us to consider the tunability characteristics (Eq. 2), i.e., the relative changes in the dielectric 'constant' due to the action of an external electric field [8, 9, 24, 28-30]:

$$T = \frac{\varepsilon(E \to 0) - \varepsilon(E)}{\varepsilon(E \to 0)} = 1 - \frac{\varepsilon(E)}{\varepsilon(E \to 0)} \qquad (38)$$

For the ferroelectric side of the para-ferro transition, where the CW Eq. (1) obeys, one obtains:

$$T = 1 - \frac{A_{CW}(E)}{A_{CW}} \frac{T - T_C}{T - T_C(E)} \qquad (39)$$

It reduces to the temperature-independent parameter $T = 1 - A_{CW}(E)/A_{CW}$ if $T_C(E)$ shift is negligible.

For the paraelectric side of the transition, related to Eq. (35), one obtains:

$$T = 1 - \frac{A}{A(E)} exp(\Delta b - \Delta a T) \qquad (40)$$

where $\Delta a = a(E) - a$ and $\Delta b = b(E) - b$, where $a$ and $b$ are related to $E = 0$.

For the 'diffused' surrounding of the para-ferro transition, one obtains:

$$T = 1 - \frac{A}{A(E)} exp(\Delta b - \Delta a T - \Delta c T^2) \qquad (41)$$

## 4. Conclusions

The report presents the model discussion of unique properties of relaxor ceramics in respect to *Critical Phenomena Physics* [47, 58, 61], *Glass Transition Physics* [39, 50, 51], and the reference to basic 'homogeneous' ferroelectrics.

It indicates the importance of pretransitional fluctuations and the essential meaning of uniaxiality for creating mean-field characterization near paraelectric – ferroelectric transition, both in 'homogeneous' and 'heterogeneous' (i.e., relaxor ceramics) materials. The discussion includes the extended Devonshire – Landau model [53, 54] and some new conclusions for the Clausius-Mossotti [81-83] local field model.

It is suggested that for the creation of characteristic $\varepsilon(T)$ changes in relaxor ceramics in the broad surrounding of the paraelectric - ferroelectric transition responsible are random local electric fields between ceramic grains with pre-ferroelectric arrangement due to pretransitional fluctuations. The impact of such local electric fields yields a distribution of local 'Curie-Weiss type' domains, associated with a set of pseudospinodal [68] singular temperatures coupled to weakly discontinuous phase transitions:

$$\varepsilon(T) = \frac{A_{Sp}^{local}}{T - T_{Sp}^{local}(E)} \qquad (42)$$

Pseudospinodal behavior leads to finite $\varepsilon(T)$ terminate values because the discontinuous transition occurs before reaching the singular temperature $T_{Sp}$. Notable, that such a picture enables avoiding problems of the essentially heuristic concept of local critical temperatures $T_C$ (Eq. 1) resulted from PNRs fluctuation, which causes local concentration changes, often recalled in modeling relaxor ceramics features [5-19, 24-26, 45, 46].

Notably, for basic, 'homogeneous', ferroelectric materials, even strong external electric field first yields non-linear changes of dielectric constant, described via so-called gap-exponents [84]. For relaxor ceramics, already moderate, external electric field strongly decreases dielectric constant ($\varepsilon'$) leading to tunability, which is crucial for applications. The given concept can be associated with the possibility of relatively easy interaction of inter-grains electric fields and the external field.

In basic 'homogeneous' ferroelectric materials, the static domain manifested via 'horizontal changes' for $\varepsilon'(f, T = const)$ scan within the frequency range $1 kHz < f < 10 MHz$ is the common feature. In such a static domain $\varepsilon'(f) \approx \varepsilon = const$, despite a frequency shift. It is also the definition of the canonic dielectric. For relaxor ceramics, such behavior is absent, and some frequency change of $\varepsilon'(f)$ in the above frequency range is a standard feature. It is shown, for example, in Figs. 3 and 4, and in the Appendix. It can also be concluded from numerous reports on relaxor ceramics. In the opinion of the authors, the frequency-dependent quasi-'dielectric constant' is the next hallmark of relaxor ceramics, a bit 'hidden' so far. Such behavior can be directly concluded from the concept-model proposed in the given report.

For the presented concept-model the spatial growth of pretransitional/pre-ferroelectric fluctuations can be estimated by the counterpart of Eq. (15a):

$$\xi(T) = \xi_0 |T - T_{Sp}(E)|^{-\nu} \qquad (43)$$

This pseudospinodal correlation length is limited by the size grain, i.e., $\xi(T) < l_{grain}$ and additionally influenced by the impact of local electric fields on the singular temperature $T_{Sp}(E)$. Such size changes are coupled to lifetime changes of fluctuations, which can be expressed by the counterpart of Eq. (15b):

$$\tau_{fl.}(T) = \tau_0 |T - T_{Sp}(E)|^{-\phi} \qquad (44)$$

Also, in the given case, terminal values are associated with the condition $\xi(T) \sim l_{grain}$. Notable that for the mean-field characterization of the system, the collective and single element relaxation time are related, i.e.: $\tau_{fl.} \propto \tau$. Hence, the distribution of grain size and the topology, as well as the impact of random local electric fields, has to yield a broad distribution of relaxation times, being the necessary prerequisite for the glassy dynamics observed in relaxor ceramics, including non-Debye and Super-Arrhenius (SA) dynamics.

It is also worth noting that the presented concept model also explains one more characteristic feature of relaxor ceramics: in different systems, the terminal values of SA change if the primary relaxation time is different, from seconds to even below milliseconds.

Experimental studies supplemented the model discussion for relaxor ceramics. The innovative distortions-sensitive and derivative-based data analysis supported them. It was possible due to the specific characteristics of the experiment and the data obtained (see Appendix).

Experimental tests were carried out in $Ba_{0.65}Sr_{0.35}TiO_3$ relaxor ceramic (see Table I). Figure 5 shows that the permanent rise of 'dielectric constant' takes place from $T \approx 120K$ to the maximum reached at $T_m \approx 291K$, and subsequently $\varepsilon(T)$ decreases down to $T_m \approx 375K$. The typical analysis applies $1/\varepsilon(T)$ vs. $T$ plot, to test the Curie-Weiss (Eq. 1) portrayal. Such plot is also shown in Figure 5, suggesting CW portrayal starting almost from $T_m \sim 292K$ to $T \approx 228K$, i.e., for $\sim 60K$. in the ferroelectric phase. In the paraelectric phase, which is the particular focus of studies recalling model analysis, the CW-type (Eq.1) behavior starts at $T_B \sim 235K$ (the Burns temperature) and terminates at $T > 375K$, i.e., for at least $\Delta T \sim 40K$. The $\Delta T = T_B - T_m$ is often considered as one of the metrics of the relaxor-type behavior, showing the width of the domain deviated from the CW behavior and linked to the appearance of Polar Nano-Regions (PNRs), hypothetically responsible for the unique behavior. For the given case $\Delta T \approx 40K$. However, the quality of experimental data enables the effective distortions-sensitive and derivative-based test of the model portrayal, avoiding a parasitic scatter often assisting in the differentiation of experimental data. Figure 6 presents such analysis focused on validating the mean-field behavior related to Curie-Weiss Eq. (1a) and Eqs. (13a) and (13b) with the exponent $\gamma = 1$. The analysis based on Eq. (27) explicitly confirms such behavior from between $T = 285K$ and $T = 234\ K$, i.e., for $\sim 50K$ in the ferroelectric phase. However, the validation for the paraelectric phase is definitely negative (!). For the paraelectric phase, the superior portrayal of $\varepsilon(T)$ changes via the exponential Eq. (30), validated by the distortions-sensitive analysis defined by Eqs. (31), (32), takes place. As shown in Figure 7, it offers a superior portrayal from 375 K to 316 K, i.e., covering ~60K (Eq. (32)). As proved by the analysis, which results presented in Figure 8 on further cooling towards the paraelectric – ferroelectric transition, the exponential relation with the additional temperature term appears. The obtained scaling patterns in the broad surrounding of the paraelectric–ferroelectric transition are concluded in Table II.

**Table II** Scaling patterns for temperature changes of 'dielectric constant' ($\varepsilon'(T)$) changes in the broad surrounding of the paraelectric – ferroelectric transition in the tested $Ba_{0.65}Sr_{0.35}TiO_3$ relaxor ceramic, specified in Table I. Note: $T_m \approx 292K$.

| Temperature range | $234K < T < 285K$ (ferro-) | $285K < T < 314K$ (para-ferro) | $315K < T < 375K$ (para-) |
|---|---|---|---|
| Scaling equation | $\varepsilon(T) = A_C/(T - T_C)$ | $\varepsilon(T) = A\exp(c + aT + bT^2)$ | $\varepsilon(T) = A\exp(b + aT)$ |

Notable is the 'negligible' distance between domains portrayed by subsequent scaling relations, the smooth passing of $T_m$ when using the (*para-ferro*) equation and the fact that the crossover from the (*para-*) to (*para-ferro*) domain is associated with the inclusion of a single, temperature-dependent term in the exponential relation.

The question arises if the obtained behavior in the (para-) and the (para-ferro) states can suggest that 'dielectric constant' changes are related to the so-called Griffiths phase [85, 86], expected for near-critical systems (particularly mean-field-type) in the presence of random impacts. For the given case, this is the randomness associated with a random local electric field between grains, which can penetrate and influence their interiors. An additional frustration can be caused by changing the properties of inter-grain layers.

The dynamic in the paraelectric phase of the tested $Ba_{0.65}Sr_{0.35}TiO_3$ system is a bit beyond the pattern observed in relaxor systems, which shows the SA-type behavior commonly portrayed by the VFT dependence. Such behavior is also observed, but it terminates at ~340K. The VFT relation can represent it, but the distortions-sensitive analysis showed the preference for activated-critical Eq. (20). For lower temperature, a new process, with explicitly Arrhenius-type temperature dependence ($E_a = const$), extending from $T \sim 330K$ to at least $T \sim 230K$. Interestingly, this unique pattern for dynamics seems to have a minimal impact of 'dielectric constant' behavior. The mentioned results have been supplemented by $tan\delta(T, f)$ behavior, focused on its physical meaning and supporting significance in testing relaxation processes: see Figs. 9, 10, 14, and the Appendix. Finally, the impact of the electric field on 'dielectric constant' was tested, revealing its strong changes for relatively moderate intensities of the electric field/voltages. For practical implementations, these features are often expressed in terms of tunability (Eq 2). The knowledge of relations describing the broad surrounding of the paraelectric – ferroelectric transitions without 'gaps' between subsequent temperature domains allowed obtaining the tunability describing dependences for the following temperature domains (Tab. II)).

In conclusion, we would like to stress that the discussion presented in this report showed the link between relaxor ceramics and *Critical Phenomena Physics* basics and *Glass transition Physics*. It indicates the meaning of uniaxiality for emerging mean-field type features. It suggests that for unique features of relaxor ceramics, particularly regarding 'dielectric constant' ($\varepsilon'(T)$) responsible can be the appearance of a random, strong, inter-grains electric field leading to the pseudospinodal behavior [68] associated with 'weakly discontinuous' phase transitions. All these re-define the meaning of the Burns temperature and Polar NanoRegions (PNRs) [3, 4], hallmark heuristic concepts recalled for explaining relaxor systems mystery [5-30].

The proposed model picture also indicates a significant impact of material engineering features on the dielectric properties of dielectric ceramics. It can be related not only to the size, composition, and structure of grains but also to the pattern of grain sintering, including relevant temperature, annealing time, and cooling/heating time rates, which can influence the growth of grains and inter-grains layers, important for the local electric field.

**Funding**
SJR, WB, MS, FG were supported by the National Science Center (NCN, Poland), project No 2018/30/Q/ST8/00205.
ADR, SS, and JŁ were supported by the NCN OPUS (Poland) grant, ref. 2022/45/B/ST5/04005.

**Author Contributions:**
Sylwester J. Rzoska (SJR) and Aleksandra Drozd-Rzoska (ADR) were responsible for the concept of the model, paper writing, and the distortions-sensitive analysis; Weronika Bulejak (WB) was responsible for samples preparation, and their non-dielectric characterization; Joanna Łoś (JŁ) carried out BDS studies and prepared a part of Figures and supported final paper preparation; Szymon Starzonek carried out measurements of dielectric constant under the strong electric field and supported the paper writing; Mikołaj Szafran (MS) and Feng Gao (FG) consulted samples preparation and their non-dielectric characterizations.
**Data Availability Statement:** Data available on reasonable requests to the corresponding author.
**Conflicts of Interest:** The authors declare no conflicts of interest with respect to the research, authorship, and/or publication of this article.

**APPENDIX**

The graphical view on the full set of obtained BDS spectra presented via frequency scans of the real ($\varepsilon'$) and imaginary ($\varepsilon''$) parts of dielectric permittivity (Fig. A1), supplemented by tan $\delta$ changes (Fig A2). Tests were carried out within ca. 200 K temperature range. For each temperature, 253 frequencies were tested, detecting impedance components with 6 digits resolution.

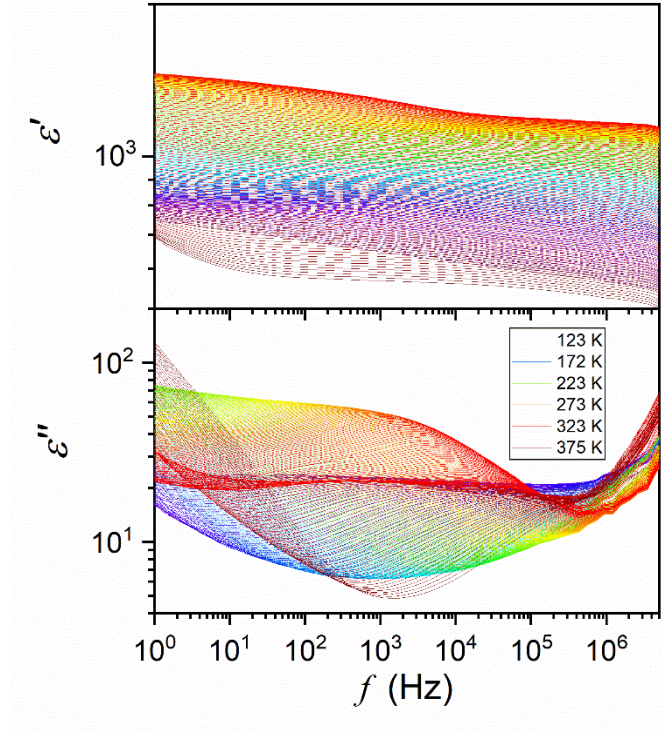

**Figure A1.** The presentation of detected complex dielectric permittivity spectra for the tests relaxor ceramic $Ba_{0.65}Sr_{0.35}TiO_3$ in the tested temperature range, indicated in the Figure.

To supplement the discussion related to $tan\delta$ in the main text of the report, we would like to stress that it is linked to energy, which can be disseminated as heat.

$$\varepsilon^* = \varepsilon' - i\varepsilon'' = \varepsilon'(1 - i \times tan\delta) \tag{A1}$$

$$tan\delta = \frac{\sigma}{\omega\varepsilon} = \frac{loss\_current}{charging\ current} \tag{A2}$$

$$D = tan\delta = \frac{i_{loss}}{i_{loss}+I} = \frac{1}{\omega RC} = \frac{\omega\varepsilon''+\sigma}{\omega\varepsilon'} = \frac{1}{Q} \tag{A3}$$

leading to the power loss:

$$P = Qtan\delta = \omega CV^2 tan\delta = \varepsilon_0 \varepsilon'' E^2 \tag{A4}$$

The latter relation, supplemented by the discussion presented in ref. [44], shows that the maximum of the loss curve ($\varepsilon''_{max}, \varepsilon''_{peak}, \varepsilon''_m$) directly express the maximal energy loss associated with the given relaxation process.

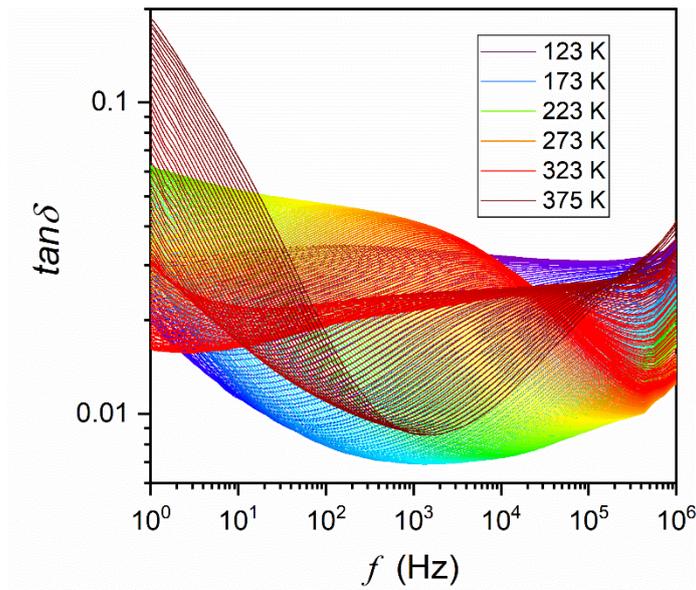

**Figure A2.** The presentation of detected BDS spectra shown via $tan\delta$ frequency scans, in the tested temperature range (indicated in the Figure) for the relaxor ceramic $Ba_{0.65}Sr_{0.35}TiO_3$, specified in Table I.